\documentclass[aps,prc,twocolumn,groupedaddress,showpacs]{revtex4-1}
\usepackage{graphicx}
\usepackage{dcolumn}
\usepackage{bm}

\begin{document}
\title{Photoelectric disintegration of $^{16}$O}
\author{M.~Katsuma}
\email{mkatsuma@sci.osaka-cu.ac.jp}
\affiliation{Advanced Mathematical Institute, Osaka City University, Osaka 558-8585, Japan}

\date{\today}

\begin{abstract}
  The photoelectric cross section of $^{16}$O($\gamma$,$\alpha$)$^{12}$C is estimated to be larger than the radiative capture cross section of $^{12}$C($\alpha$,$\gamma$)$^{16}$O.
  The predicted cross section and the angular distribution of $\alpha$-particle are illustrated for the future experiment.
  The cross section just above the $\alpha$-particle threshold is found to be dominated by the $E$2 excitation.
\end{abstract}


\maketitle

  The low-energy $^{12}$C($\alpha$,$\gamma$)$^{16}$O reaction plays the crucial role in the nucleosynthesis of element in a star.
  However, the cross section is very small at the energies corresponding to helium burning temperatures, $E_{c.m.}\approx 0.3$ MeV, because of the Coulomb barrier, and it is far beyond out of reach by the present laboratory technologies \cite{Rol88}.
  ($E_{c.m.}$ is the center-of-mass energy of the $\alpha$+$^{12}$C system.)
  To cope with the difficulties, the experimental challenges have continuously been developed (e.g. \cite{Gai14,Gai13,Bru13,Sch12,Pla12,Mak09,Ass06,Kun01}), as well as the theoretical predictions \cite{Kat08,Kat14b}.
  The current experimental projects \cite{Gai14,Ahm13} of photonuclear reactions with high intensity laser are the one of the precise measurements for the tiny cross section.

  The probability of the photodisintegration of $^{16}$O just above the $\alpha$-particle threshold restricts the reaction rates of the $^{12}$C($\alpha$,$\gamma$)$^{16}$O reaction.
  In particular, the angular distribution of the emitted $\alpha$-particle is expected to reduce the large uncertainties of the cross section.
  The subthreshold 1$^-_1$ state at the excitation energy $E_x=7.12$ MeV is believed to couple strongly with the 1$^-_2$ state at $E_x=9.585$ MeV.
  The strong interference with two 1$^-$ states lead to the large enhancement of the low-energy cross section, and it has been presumed to play the important role in the derived reaction rates at helium burning temperatures.
  In contrast, the $E$1 cross section is predicted not to be enhanced by the high energy tail of the subthreshold 1$^-_1$ state because the $\alpha$+$^{12}$C system can be described by the weak coupling \cite{Kat08,Kat14b,Kat10b}.
  In our previous studies, the low-energy cross section is dominated by the $E$2 transition.
  This has been endorsed by the $\gamma$-ray angular distribution at $E_{c.m.}=1.254$ -- 1.34 MeV \cite{Kun01,Ass06} and the transparency of the $\alpha$+$^{12}$C system at low energies \cite{Kat10b,Kat14b,Tis09,Pla87}.

  In the present report, we study the photoelectric disintegration of $^{16}$O.
  Using the reciprocity theorem, we illustrate the expected photoelectric cross section of the $^{16}$O($\gamma$,$\alpha$)$^{12}$C reaction.

  Let us recall the relation between the photonuclear reaction and the capture reaction.
  The cross section of the $^{16}$O($\gamma$,$\alpha$)$^{12}$C reaction is given by the inverse reaction in the following expression,
  \begin{eqnarray}
    \sigma_{\gamma\alpha} (E_\gamma) &=& \frac{k_{\rm c}^2}{2k_\gamma^2} \sigma_{\alpha\gamma} (E_{c.m.})
    \label{eq:sig}
  \end{eqnarray}
  where $k_{\rm c}$ is the wavenumber of relative motion between $\alpha$-particle and $^{12}$C nuclei; 
  $k_\gamma$ is the wavenumber of photon $k_\gamma=E_\gamma/(\hbar c)$, $E_\gamma=E_{c.m.}+7.162$ MeV;
  $\sigma_{\alpha\gamma}$ is the capture cross section for the $^{12}$C($\alpha$,$\gamma$)$^{16}$O reaction.
  To obtain the predicted values, we use the same potential model as \cite{Kat08}.

  Owing to the Coulomb barrier of the $\alpha$+$^{12}$C channel, the $\alpha$-particle cannot be emitted easily above the threshold energy.
  The resulting cross section is very small.
  To compensate for the rapid energy variation, we illustrate the cross section multiplied by the Gamow factor, $\exp(2\pi\eta)$,  in the present report.
  $\eta$ is the Sommerfeld parameter, $\eta=Z_1Z_2 e^2/(\hbar v)$.
  $v$ is the velocity of relative motion between $\alpha$-particle and $^{12}$C nuclei.
  $Z_1$ and $Z_2$ are the charges of the interacting nuclei.

  The cross section of the photodisintegration is expected to be larger than that of the $^{12}$C($\alpha$,$\gamma$)$^{16}$O reaction.
  From Eq.~(\ref{eq:sig}), we find $\sigma_{\gamma\alpha}\approx 50 \times \sigma_{\alpha\gamma}$ at $E_\gamma\approx 8.41$ MeV, corresponding to $E_{c.m.}= 1.25$ MeV close to the current lowest energy of the angular distribution measurement of $^{12}$C($\alpha$,$\gamma$)$^{16}$O \cite{Kun01}.
  This makes it possible to determine the $E$2/$E$1 ratio of the cross section more accurately.

\begin{figure}[t]
  \includegraphics[width=0.98\linewidth]{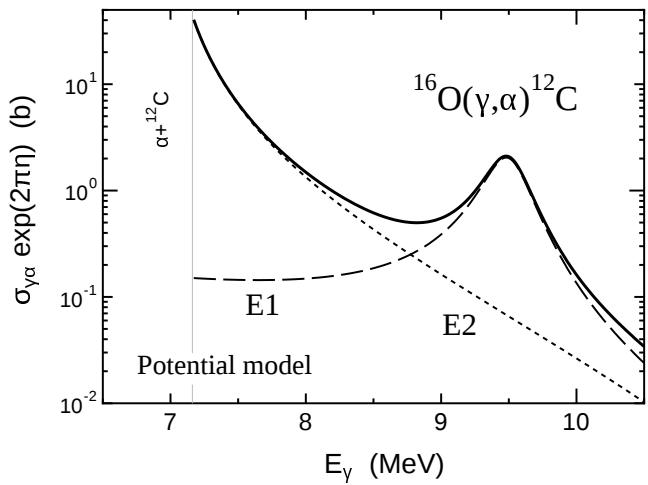}
  \caption{\label{fig:sig}
    Photoelectric cross section for the $^{16}$O($\gamma$,$\alpha$)$^{12}$C reaction as a function of the $\gamma$-ray energy.
    The solid curve is the result obtained from the potential model.
    The dashed and dotted curves are the $E$1 and $E$2 components, respectively.
    The cross section is multiplied by $\exp(2\pi\eta)$ to compensate for the rapid energy variation in the vicinity of the $\alpha$-particle threshold.
    The vertical thin line at $E_\gamma=7.162$ MeV indicates the energy position of the threshold.
  }
\end{figure}
\begin{figure}[t]
  \includegraphics[width=0.85\linewidth]{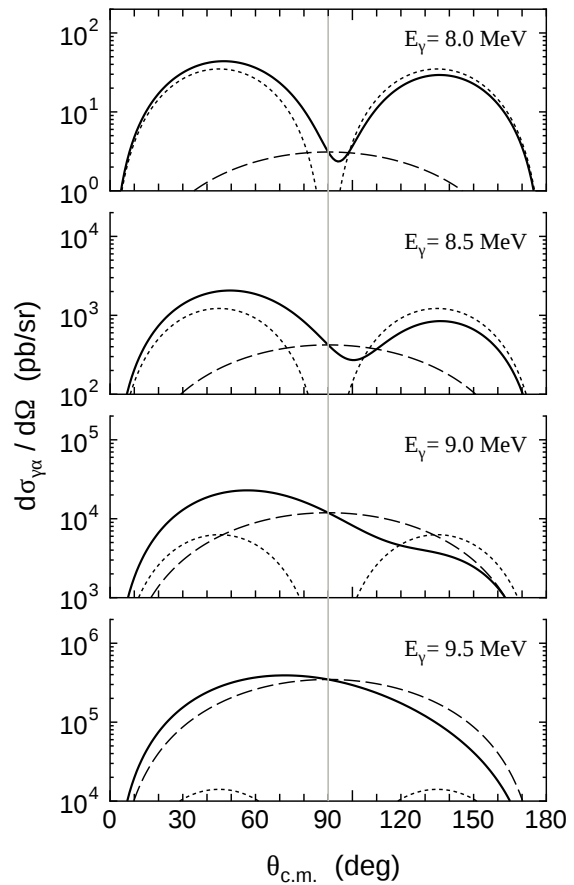}
  \caption{\label{fig:ang}
    Differential cross sections for the $^{16}$O($\gamma$,$\alpha$)$^{12}$C reaction at $E_\gamma=8.0$, 8.5, 9.0 and 9.5 MeV.
    The solid curves are the results obtained from the potential model.
    The dashed and dotted curves are the pure $E$1 and $E$2 components.
  }
\end{figure}

  In Fig.~\ref{fig:sig}, we illustrate the calculated photoelectric cross section of the $^{16}$O($\gamma$,$\alpha$)$^{12}$C reaction.
  The solid curve is the result obtained from the potential model.
  The dashed and dotted curves are the $E$1 and $E$2 components, respectively.
  The vertical thin line at $E_\gamma=7.162$ MeV indicates the energy position of the $\alpha$-particle threshold in $^{16}$O.
  The $\alpha$-particle is emitted above this energy.
  The peak at the 1$^-_2$ state can be seen at $E_\gamma\approx9.5$ MeV.
  The $E$1 component is approximately constant below $E_\gamma= 8$ MeV.
  The $E$2 cross section is enhanced as $E_\gamma$ decreases.
  The photoelectric disintegration is found to be dominated by the $E$2 excitation below $E_\gamma\approx 8$ MeV.
  This is due to the high energy tail of the subthreshold 2$^+_1$ state at $E_x=6.92$ MeV, which has the well-developed $\alpha$+$^{12}$C cluster structure \cite{Kat10b}.

  The $\sigma_{\gamma\alpha}\exp(2\pi\eta)$ in Fig.~\ref{fig:sig} appears to resemble the astrophysical $S$-factor in the $^{12}$C($\alpha$,$\gamma$)$^{16}$O reaction \cite{Kat08}.
  The energy variation, in fact, corresponds to $\sigma_{\gamma\alpha}\exp(2\pi\eta)\propto S/E_\gamma^2$.

  The photonuclear reaction may remind us of the discussion about the dipole excitation.
  However, the photomagnetic dipole excitation ($s$-wave) is forbidden.
  The electric dipole transition ($p$-wave) is hindered by the isospin selection rule.
  Thus, the $E$2 excitation ($d$-wave) could be dominant mode of the transition in the vicinity of the $\alpha$-particle threshold.
  The $E$3 and $E$4 transitions ($f$- and $g$-waves) are negligible.

  The angular distributions of the $\alpha$-particle at $E_\gamma=8.0$, 8.5, 9.0 and 9.5 MeV are shown in Fig.~\ref{fig:ang}.
  The solid curves are the results obtained from the potential model.
  $\theta_{c.m.}$ is defined with respect to the beam direction.
  At $E_\gamma=9.5$ MeV, the angular distribution appear to be the single peak due to the 1$^-_2$ state at $E_x=9.585$ MeV.
  The angular distribution becomes the double peaks at $E_\gamma=8.0$ MeV.
  The dashed and dotted curves are the pure $E$1 and $E$2 components.
  The $E$2 contribution caused by the subthreshold 2$^+_1$ state interferes with the dipole component.

  The differential cross section at $\theta_{c.m.}=90^\circ$ can be made only from the $E$1 component, because the $E$2 component vanishes at this angle.
  Using the $90^\circ$ cross section, the integrated $E$1 cross section is expressed as
  \begin{eqnarray}
    \sigma^{E1}_{\gamma\alpha} (E_\gamma) &=&  \frac{8\pi}{3} \left. \frac{d\sigma_{\gamma\alpha}}{d\Omega} \right|_{\theta_{c.m.}=90^\circ}.
    \label{eq:sig90}
  \end{eqnarray}
  The dashed curve in Fig.~\ref{fig:sig} can be confirmed from this relation.
  The numerical values are listed in Table~\ref{tb:sig}.

\begin{table}[t]
  \caption{\label{tb:sig}
    Photoelectric dipole cross section and $90^\circ$ cross section.
    The value of $\exp(2\pi\eta)$ is also listed.
  }
  \begin{ruledtabular}
  \begin{tabular}{cccc}
    $E_\gamma$ (MeV) & d$\sigma_{\gamma\alpha}$/d$\Omega$(90$^\circ$) (b/sr) & $\sigma^{E1}_{\gamma\alpha}$ (b) & $\exp(2\pi\eta)$ \\
    \hline
    8.0 & 3.11$\times10^{-12}$ & 2.60$\times10^{-11}$ & 5.72$\times10^9$ \\
    8.5 & 4.22$\times10^{-10}$ & 3.53$\times10^{-9}$ & 5.27$\times10^7$\\
    9.0 & 1.19$\times10^{-8}$ & 9.98$\times10^{-8}$ & 3.88$\times10^6$\\
    9.5 & 3.47$\times10^{-7}$ & 2.91$\times10^{-6}$ & 6.94$\times10^5$\\
  \end{tabular}
  \end{ruledtabular}
\end{table}

  The 90$^\circ$ cross section basically determines the $E$1 contribution of the photodisintegration of $^{16}$O.
  In addition, the $E$2 contribution is predicted to be large at low energies, so the angular distribution has the minimum at $\theta_{c.m.}\approx90^\circ$.
  In this circumstance, the $E$1 component may be susceptible to the background noise.
  The absolute value of the $E$1 cross section could not be determined without precise measurement of the angular distribution.
  We expect that the $E$2/$E$1 ratio will be provided more accurately by the future experiment.

  In summary, the $^{16}$O($\gamma$,$\alpha$)$^{12}$C reaction has been studied with the potential model.
  We have shown the calculated photoelectric cross section and the angular distribution of $\alpha$-particle.
  The cross section just above the $\alpha$-particle threshold is dominated by the $E$2 excitation.
  The cross section of the photoelectric disintegration is found to be larger than that of the $^{12}$C($\alpha$,$\gamma$)$^{16}$O reaction.
  We expect that the forthcoming experimental projects \cite{Gai14,Ahm13} will determine the $^{12}$C($\alpha$,$\gamma$)$^{16}$O cross section more accurately.

  The author is grateful to Prof.~Moshe Gai and Prof. M. Fujiwara for their encouragement.
  He thanks Y.~Kond\=o for early days of collaboration and Y. Ohnita and Y.~Sakuragi for their hospitality.
  He also thanks M.~Arnould, A.~Jorissen, K.~Takahashi, and H.~Utsunomiya for their hospitality at his stay in Universit\'e Libre de Bruxelles.


\begin{thebibliography}{19}
\expandafter\ifx\csname natexlab\endcsname\relax\def\natexlab#1{#1}\fi
\expandafter\ifx\csname bibnamefont\endcsname\relax
  \def\bibnamefont#1{#1}\fi
\expandafter\ifx\csname bibfnamefont\endcsname\relax
  \def\bibfnamefont#1{#1}\fi
\expandafter\ifx\csname citenamefont\endcsname\relax
  \def\citenamefont#1{#1}\fi
\expandafter\ifx\csname url\endcsname\relax
  \def\url#1{\texttt{#1}}\fi
\expandafter\ifx\csname urlprefix\endcsname\relax\def\urlprefix{URL }\fi
\providecommand{\bibinfo}[2]{#2}
\providecommand{\eprint}[2][]{\url{#2}}

\bibitem[{\citenamefont{Rolfs and Rodney}(1988)}]{Rol88}
\bibinfo{author}{\bibfnamefont{C.~E.} \bibnamefont{Rolfs}} \bibnamefont{and}
  \bibinfo{author}{\bibfnamefont{W.~S.} \bibnamefont{Rodney}},
  \emph{\bibinfo{title}{Cauldrons in the Cosmos}} (\bibinfo{publisher}{The
  University of Chicago Press}, \bibinfo{year}{1988}).

\bibitem[{\citenamefont{Gai}(2014)}]{Gai14}
\bibinfo{author}{\bibfnamefont{M.}~\bibnamefont{Gai}}, \bibinfo{journal}{Nucl. Phys.}
  \textbf{\bibinfo{volume}{A928}}, \bibinfo{pages}{313} (\bibinfo{year}{2014}).

\bibitem[{\citenamefont{Gai}(2013)}]{Gai13}
\bibinfo{author}{\bibfnamefont{M.}~\bibnamefont{Gai}}, \bibinfo{journal}{Phys.
  Rev. C} \textbf{\bibinfo{volume}{88}}, \bibinfo{pages}{062801}
  (\bibinfo{year}{2013}).

\bibitem[{\citenamefont{Brune}(2013)}]{Bru13}
\bibinfo{author}{\bibfnamefont{C.}~\bibnamefont{Brune}}, \bibinfo{journal}{J.
  Phys. Conf.} \textbf{\bibinfo{volume}{420}}, \bibinfo{pages}{012140}
  (\bibinfo{year}{2013}).

\bibitem[{\citenamefont{Sch\"urmann et~al.}(2012)\citenamefont{Sch\"urmann,
  Gialanella, Kunz, and Strieder}}]{Sch12}
\bibinfo{author}{\bibfnamefont{D.}~\bibnamefont{Sch\"urmann}},
  \bibinfo{author}{\bibfnamefont{L.}~\bibnamefont{Gialanella}},
  \bibinfo{author}{\bibfnamefont{R.}~\bibnamefont{Kunz}}, \bibnamefont{and}
  \bibinfo{author}{\bibfnamefont{F.}~\bibnamefont{Strieder}},
  \bibinfo{journal}{Phys. Lett.} \textbf{\bibinfo{volume}{B711}},
  \bibinfo{pages}{35} (\bibinfo{year}{2012}).

\bibitem[{\citenamefont{Plag et~al.}(2012)\citenamefont{Plag, Reifarth, Heil,
  K\"appeler, Rupp, Voss, and Wisshak}}]{Pla12}
\bibinfo{author}{\bibfnamefont{R.}~\bibnamefont{Plag}},
  \bibinfo{author}{\bibfnamefont{R.}~\bibnamefont{Reifarth}},
  \bibinfo{author}{\bibfnamefont{M.}~\bibnamefont{Heil}},
  \bibinfo{author}{\bibfnamefont{F.}~\bibnamefont{K\"appeler}},
  \bibinfo{author}{\bibfnamefont{G.}~\bibnamefont{Rupp}},
  \bibinfo{author}{\bibfnamefont{F.}~\bibnamefont{Voss}}, \bibnamefont{and}
  \bibinfo{author}{\bibfnamefont{K.}~\bibnamefont{Wisshak}},
  \bibinfo{journal}{Phys. Rev. C} \textbf{\bibinfo{volume}{86}},
  \bibinfo{pages}{015805} (\bibinfo{year}{2012}).

\bibitem[{\citenamefont{Makii et~al.}(2009)\citenamefont{Makii, Nagai, Shima,
  Segawa, Mishima, Ueda, Igashira, and Ohsaki}}]{Mak09}
\bibinfo{author}{\bibfnamefont{H.}~\bibnamefont{Makii}},
  \bibinfo{author}{\bibfnamefont{Y.}~\bibnamefont{Nagai}},
  \bibinfo{author}{\bibfnamefont{T.}~\bibnamefont{Shima}},
  \bibinfo{author}{\bibfnamefont{M.}~\bibnamefont{Segawa}},
  \bibinfo{author}{\bibfnamefont{K.}~\bibnamefont{Mishima}},
  \bibinfo{author}{\bibfnamefont{H.}~\bibnamefont{Ueda}},
  \bibinfo{author}{\bibfnamefont{M.}~\bibnamefont{Igashira}}, \bibnamefont{and}
  \bibinfo{author}{\bibfnamefont{T.}~\bibnamefont{Ohsaki}},
  \bibinfo{journal}{Phys. Rev. C} \textbf{\bibinfo{volume}{80}},
  \bibinfo{pages}{065802} (\bibinfo{year}{2009}).

\bibitem[{\citenamefont{Assun\c~c\ ao et~al.}(2006)\citenamefont{Assun\c~c\ ao,
  Fey, Lefebvre-Schuhl, Kiener, Tatischeff, Hammer, Beck, Boukari-Pelissie,
  Coc, Correia et~al.}}]{Ass06}
\bibinfo{author}{\bibfnamefont{M.}~\bibnamefont{Assun\c c\~ao}},
  \bibinfo{author}{\bibfnamefont{M.}~\bibnamefont{Fey}},
  \bibinfo{author}{\bibfnamefont{A.}~\bibnamefont{Lefebvre-Schuhl}},
  \bibinfo{author}{\bibfnamefont{J.}~\bibnamefont{Kiener}},
  \bibinfo{author}{\bibfnamefont{V.}~\bibnamefont{Tatischeff}},
  \bibinfo{author}{\bibfnamefont{J.~W.} \bibnamefont{Hammer}},
  \bibinfo{author}{\bibfnamefont{C.}~\bibnamefont{Beck}},
  \bibinfo{author}{\bibfnamefont{C.}~\bibnamefont{Boukari-Pelissie}},
  \bibinfo{author}{\bibfnamefont{A.}~\bibnamefont{Coc}},
  \bibinfo{author}{\bibfnamefont{J.~J.} \bibnamefont{Correia}},
  \bibnamefont{et~al.}, \bibinfo{journal}{Phys. Rev. C}
  \textbf{\bibinfo{volume}{73}}, \bibinfo{pages}{055801}
  (\bibinfo{year}{2006}).

\bibitem[{\citenamefont{Kunz et~al.}(2001)\citenamefont{Kunz, Jaeger, Mayer,
  Hammer, Staudt, Harissopulos, and Paradellis}}]{Kun01}
\bibinfo{author}{\bibfnamefont{R.}~\bibnamefont{Kunz}},
  \bibinfo{author}{\bibfnamefont{M.}~\bibnamefont{Jaeger}},
  \bibinfo{author}{\bibfnamefont{A.}~\bibnamefont{Mayer}},
  \bibinfo{author}{\bibfnamefont{J.~W.} \bibnamefont{Hammer}},
  \bibinfo{author}{\bibfnamefont{G.}~\bibnamefont{Staudt}},
  \bibinfo{author}{\bibfnamefont{S.}~\bibnamefont{Harissopulos}},
  \bibnamefont{and}
  \bibinfo{author}{\bibfnamefont{T.}~\bibnamefont{Paradellis}},
  \bibinfo{journal}{Phys. Rev. Lett.} \textbf{\bibinfo{volume}{86}},
  \bibinfo{pages}{3244} (\bibinfo{year}{2001}).

\bibitem[{\citenamefont{Katsuma}(2008)}]{Kat08}
\bibinfo{author}{\bibfnamefont{M.}~\bibnamefont{Katsuma}},
  \bibinfo{journal}{Phys. Rev. C} \textbf{\bibinfo{volume}{78}},
  \bibinfo{pages}{034606} (\bibinfo{year}{2008});
  \bibinfo{journal}{ibid.} \textbf{\bibinfo{volume}{81}},
  \bibinfo{pages}{029804} (\bibinfo{year}{2010});
  \bibinfo{journal}{Astrophys. J.} \textbf{\bibinfo{volume}{745}},
  \bibinfo{pages}{192} (\bibinfo{year}{2012}).

\bibitem[{\citenamefont{Katsuma}(2014{\natexlab{a}})}]{Kat14b}
\bibinfo{author}{\bibfnamefont{M.}~\bibnamefont{Katsuma}},
  \bibinfo{journal}{arXiv:}\bibinfo{pages}{1404.3966}
  (\bibinfo{year}{2014}{\natexlab{a}}).

\bibitem[{\citenamefont{Ahmed et~al.}(2013)\citenamefont{Ahmed, Champagne,
  Holstein, Howell, Snow, Springer, and Wu}}]{Ahm13}
\bibinfo{author}{\bibfnamefont{M.}~\bibnamefont{Ahmed}},
  \bibinfo{author}{\bibfnamefont{A.}~\bibnamefont{Champagne}},
  \bibinfo{author}{\bibfnamefont{B.}~\bibnamefont{Holstein}},
  \bibinfo{author}{\bibfnamefont{C.}~\bibnamefont{Howell}},
  \bibinfo{author}{\bibfnamefont{W.}~\bibnamefont{Snow}},
  \bibinfo{author}{\bibfnamefont{R.}~\bibnamefont{Springer}}, \bibnamefont{and}
  \bibinfo{author}{\bibfnamefont{Y.}~\bibnamefont{Wu}},
  \bibinfo{journal}{arXiv:}\bibinfo{pages}{1307.8178}
  (\bibinfo{year}{2013}).

\bibitem[{\citenamefont{Katsuma}(2010{\natexlab{b}})}]{Kat10b}
\bibinfo{author}{\bibfnamefont{M.}~\bibnamefont{Katsuma}},
  \bibinfo{journal}{Phys. Rev. C} \textbf{\bibinfo{volume}{81}},
  \bibinfo{pages}{067603} (\bibinfo{year}{2010});
  \bibinfo{journal}{J. Phys. G} \textbf{\bibinfo{volume}{40}}, \bibinfo{pages}{025107} (\bibinfo{year}{2013});
  \bibinfo{journal}{Proceedings of INPC2013, EPJ Web of Conferences} \textbf{\bibinfo{volume}{66}}, \bibinfo{pages}{03041} (\bibinfo{year}{2014}).

\bibitem[{\citenamefont{Tischhauser et~al.}(2009)\citenamefont{Tischhauser,
  Couture, Detwiler, G\"orres, Ugalde, Stech, Wiescher, Heil, K\"appeler, Azuma
  et~al.}}]{Tis09}
\bibinfo{author}{\bibfnamefont{P.}~\bibnamefont{Tischhauser}},
  \bibinfo{author}{\bibfnamefont{A.}~\bibnamefont{Couture}},
  \bibinfo{author}{\bibfnamefont{R.}~\bibnamefont{Detwiler}},
  \bibinfo{author}{\bibfnamefont{J.}~\bibnamefont{G\"orres}},
  \bibinfo{author}{\bibfnamefont{C.}~\bibnamefont{Ugalde}},
  \bibinfo{author}{\bibfnamefont{E.}~\bibnamefont{Stech}},
  \bibinfo{author}{\bibfnamefont{M.}~\bibnamefont{Wiescher}},
  \bibinfo{author}{\bibfnamefont{M.}~\bibnamefont{Heil}},
  \bibinfo{author}{\bibfnamefont{F.}~\bibnamefont{K\"appeler}},
  \bibinfo{author}{\bibfnamefont{R.~E.} \bibnamefont{Azuma}},
  \bibnamefont{et~al.}, \bibinfo{journal}{Phys. Rev. C}
  \textbf{\bibinfo{volume}{79}}, \bibinfo{pages}{055803}
  (\bibinfo{year}{2009}).

\bibitem[{\citenamefont{Plaga et~al.}(1987)\citenamefont{Plaga, Becker, Redder,
  Rolfs, Trautvetter, and Langanke}}]{Pla87}
\bibinfo{author}{\bibfnamefont{R.}~\bibnamefont{Plaga}},
  \bibinfo{author}{\bibfnamefont{H.~W.} \bibnamefont{Becker}},
  \bibinfo{author}{\bibfnamefont{A.}~\bibnamefont{Redder}},
  \bibinfo{author}{\bibfnamefont{C.}~\bibnamefont{Rolfs}},
  \bibinfo{author}{\bibfnamefont{H.~P.} \bibnamefont{Trautvetter}},
  \bibnamefont{and} \bibinfo{author}{\bibfnamefont{K.}~\bibnamefont{Langanke}},
  \bibinfo{journal}{Nucl. Phys.} \textbf{\bibinfo{volume}{A465}},
  \bibinfo{pages}{291} (\bibinfo{year}{1987}).

\end{thebibliography}

\end{document}